# Automated Inline Analysis of Myocardial Perfusion MRI with Deep Learning

Hui Xue[1], Rhodri Davies[2], Louis AE Brown[3], Kristopher D Knott[2], Tushar Kotecha[4], Marianna Fontana[4], Sven Plein[3], James C Moon[2], Peter Kellman[1]

1. National Heart, Lung and Blood Institute, National Institutes of Health, Bethesda, MD, USA
2. Barts Heart Centre, Barts Health NHS Trust, London, UK
3. Department of Biomedical Imaging Science, Leeds Institute of Cardiovascular and Metabolic Medicine, University of Leeds, Leeds, UK
4. National Amyloidosis Centre, Royal Free Hospital, London, UK

## Corresponding author:

Hui Xue

National Heart, Lung and Blood Institute
National Institutes of Health
10 Center Drive, Bethesda
MD 20892
USA

Phone: +1 (301) 827-0156
Cell:    +1 (609) 712-3398
Fax:     +1 (301) 496-2389
Email: hui.xue@nih.gov

Word Count: 2,000

hui.xue@nih.gov
rhodri.davies2@nhs.net
kristopher.knott@nhs.net
l.brown1@leeds.ac.uk
s.plein@leeds.ac.uk
tushar.kotecha@nhs.net
m.fontana@ucl.ac.uk
j.moon@ucl.ac.uk
kellmanp@nhlbi.nih.gov

# Automated Inline Analysis of Myocardial Perfusion MRI with Deep Learning

**Author list**

Author affiliations

**Key points:**

1. Proposed and validated a convolutional neural network solution for cardiac perfusion mapping and integrated an automated inline implementation on the MR scanner, enabling "one-click" analysis and reporting.

2. Large training set included N=1,825 perfusion series from 1,034 patients. Independent test set included 200 scans from 105 patients.

3. Comparison of automated and manual derived myocardial blood flow measurement showed no statistically differences on both global and per-sector basis (P>0.80).

**Summary statement:**

This paper proposed, validated, and deployed a convolutional neural network solution for automated analysis of cardiac perfusion MR imaging.

**Word Count: 2,000**

**Abstract**

**Background**

Pixel-wise quantitative myocardial blood flow (MBF) mapping allows objective assessment of ischemic heart disease. Objective assessment can be further facilitated by automatically segmenting the myocardium and generating reports following the AHA model. This paper proposes a deep neural network based computational workflow for inline myocardial perfusion analysis that automatically delineates the myocardium, which improves the clinical workflow and leads to a "one-click" solution.

**Methods**

Consecutive, adenosine stress/rest perfusion scans were acquired from three hospitals. The training set included N=1,825 perfusion series from 1,034 patients (mean age 60.6±14.2yrs). The independent test set included 200 scans from 105 patients (mean age 59.1±12.5 yrs). A convolutional neural network (CNN) model was trained to segment the LV cavity, myocardium and right ventricle by processing an incoming time series of perfusion images (2D+T). Model outputs were compared to manual ground-truth for accuracy of segmentation and flow measures derived on a global and per-sector basis. The trained models were integrated onto MR scanners for effective inference.

**Results**

Mean Dice ratio of automatic and manual segmentation was 0.93 ± 0.04. Mean stress MBF (ml/min/g) was 2.25±0.59 (CNN) and 2.24±0.59 (manual) (P=0.94). Mean rest MBF was 1.08±0.23 (CNN) and 1.07±0.23 (manual) (P=0.83). The per-sector MBF values showed no significant difference (P=0.92). CPU based model inference on the MR scanner took <1s for a typical perfusion scan of three slices.

**Conclusions**

This study proposed and validated a convolutional neural network solution for cardiac perfusion mapping and integrated an automated inline implementation on the MR scanner, enabling "one-click" analysis and reporting.

**Key words**

# Introduction

Myocardial perfusion MRI has proven to be an accurate non-invasive imaging technique to detect ischemic heart disease (1). Quantitative MR perfusion is more objective (2–5) and automated in-line methods (6–8) offer improved efficiency of analysis. Compared with visual assessment, quantitative methods improve the detection of disease with a global reduction in flow, as seen in balanced multi-vessel obstruction or microvascular disease (4,5,9).

Without automated segmentation of the MR perfusion maps, a reporting clinician would have to manually draw regions of interest to extract global or regional flow values. Objective perfusion assessment can be further facilitated by segmenting the myocardium to automatically generate the report leading to a "one-click" solution to improve workflow. Automated MR perfusion measurement could serve as the input for down-stream cardiovascular disease classification (10–12) where pre-trained CNN models receive myocardial flow and other biomarkers to predict the probability of ischemic heart disease. These studies used manual segmentation and can be automated with the proposed approach.

In this study we propose a deep CNN based computational workflow for myocardial perfusion analysis using MRI. The right ventricular (RV) insertion points were determined to allow reporting of perfusion according to the standard 16 segment model proposed by the American Heart Association (AHA). To utilize the dynamic change of intensity due to contrast uptake, the proposed solution operates on the time series of perfusion images (referred to here as 2D+T) after respiratory motion correction. The performance of trained CNNs was quantitatively evaluated by comparing against manually established ground-truth for both segmentation accuracy and global as well as regional flow measures on an independent hold-out test dataset.

To promote the clinical validation and adoption of the proposed solution, the trained deep learning models were integrated onto MR scanners using the Gadgetron InlineAI toolbox (13). The CNN models were applied to the acquired images as part of the scanner computing workflow (inline processing) at the time of scan, rather than as a post processing. The resulting segmentation results and analysis reports were available for immediate evaluation prior to the next image series. The method described here has been used in a prospective study of >1000 patients to demonstrate the prognostic significance of quantitative stress perfusion (2). A "one-click" solution to acquire free-breathing perfusion images, perform pixel-wise flow mapping and conduct automated analysis with a 16-segment AHA report generated on the MR scanner is demonstrated.

## Methods

*Imaging and data collection*

The dataset consists of adenosine stress and rest perfusion scans acquired at three hospitals (Barts Heart Centre, Barts; Royal Free Hospital, RFH; Leeds Teaching Hospitals NHS Trust, LTHT). N=1,825 perfusion scans (791 patients had both stress and rest perfusion; others had rest perfusion) from 1,034 patients (mean age 60.6±14.2 yrs) were assembled and split into training and validation sets, used for CNN model training. Training and validation data were consecutively acquired at each site (Barts: 475 patients; RFH: 345 patients; LTHT: 214 patients), including 791 stress and 1,034 rest scans. An independent hold-out consecutive test set was assembled, consisting of 200 perfusion scans (95 patients with stress and rest scans) from 105 patients (mean age 59.1±12.5 yrs; Barts: 54 patients; RFH: 13 patients; LTHT: 38 patients), including 95 stress and 105 rest scans. There was no overlap between training/validation data and independent test data (14). Among the assembled independent test data, 96 scans were acquired at 3T scanners and 104 were from 1.5T scanners.

Perfusion imaging used a previously published "dual-sequence" scheme (6). A low-resolution arterial input function (AIF) imaging module was inserted before the perfusion imaging and performed after the R-wave with short delay time. For stress perfusion, adenosine was administered by continuous intravenous infusion for 4 min at a dose of 140 µg/kg/min before contrast injection (increased to 175 µg/kg per minute for a further 2 minutes based on patient's response). The imaging started by acquiring 3 proton density weighted (PD) images, followed by saturation recovery images. Every perfusion image was acquired as a 2D image cutting through the heart and this acquisition was repeated for every heart beat to capture the contrast passage, typically lasting 60 heart beats. This resulted in the 2D+T time series where images were acquired consecutively in time. Details of imaging and perfusion mapping can be found in (6,7). Datasets were acquired using both 1.5T (four MAGNETOM Aera, Siemens AG Healthcare, Erlangen, Germany) and 3T (three MAGNETOM Prisma, Siemens AG Healthcare) MR scanners.

Data was acquired with the required ethical and/or audit secondary use approvals or guidelines (as per each center) that permitted retrospective analysis of anonymized data for the purpose of technical development, protocol optimization and quality control. All data was anonymized and de-linked for analysis by NIH with approval by the NIH Office of Human Subjects Research OHSR (Exemption #13156).

*Data preparation and labeling*

Perfusion image series underwent motion correction and surface coil inhomogeneity correction. Motion correction utilized non-rigid image registration in an iterative manner. To cope with significant image contrast variation during the contrast bolus passage, instead of directly registering perfusion images against each other, synthetic perfusion series were derived from a Karhunen-Loève transform. Motion correction was achieved by registering perfusion images pairwise with the synthetic series. The detailed algorithm was presented in (6,7). After correcting respiratory motion, surface coil inhomogeneity was corrected using the proton density images and the normalized intensities were converted to [Gd] concentration units (mmol/L) (6,7). To compensate for heart rate variation and mis-triggering, the perfusion series was temporally resampled using linear interpolation which also compensated for possible missed triggers. This interpolation resulted in a fixed sampling corresponding to a heart rate (HR) of 120 bpm. The temporal resampling step did not lead to spatial blurring since it was performed after motion correction.

Since the Gd concentration series was corrected for signal nonlinearity and surface coil inhomogeneity, it had the benefit of reducing the dynamic range and providing a fixed signal range for neural nets, compared to perfusion intensity images. This image series was spatially upsampled to $1.0mm^2$ spatial resolution and the central FOV ($176 \times 176mm^2$) was cropped. The LV blood pool was detected from the AIF series which was imaged at the basal plane at diastole. The location of the LV blood pool from this step was used to center the cropped image (15). For the SAX perfusion slices, the LV endo- and epicardial boundaries were manually traced, together with the right ventricle (RV) (Fig. 1). The RV insertion point (RVI) was determined from the segmented right ventricular and LV center as the rightmost pixel. Training and test datasets were carefully labeled by one operator (HX, 10 years of experience in perfusion imaging).

***Neural Net model***

The first 48 images were empirically selected, starting at the first saturation recovery image. This resulted in an image array of 176×176×48 per slice, covering the first-pass bolus passage of injected contrast agent. A total of 262,800 2D images were then used for training the neural networks. The U-net semantic segmentation architecture (16,17) was adopted for the perfusion segmentation. The neural net (Fig. 2) consisted of downsampling and upsampling layers, each including a number of ResNet blocks (18). The downsampling and upsampling operations were inserted between layers to change the spatial resolution. For simplicity, two convolutional layer operations with the same number of output filters were added to each block, together with Batch Normalization (BN) (19) and LeakyRelu (20) nonlinearity. All convolutional layers used a 3x3 kernel with stride 1 and padding 1. Following the principle of U-net, the downsampling and upsampling layers were connected with "Skip-connections". The spatial resolution was reduced by going through the down-sampling branch with the number of convolution filters increased. The up-sampling branch increased the spatial resolution and reduced the number of filters. The network was able to learn features from this coarse-to-fine pyramid thereby selecting an optimal filter combination to minimize the loss function.

The final convolutional layer output was a 176×176×3 array of scores representing segmented classes, which were converted to probability through a softmax operation. Establishing the anatomical context of LV cavity, myocardium and RV was facilitated by using a single trained CNN. The loss function was a weighted sum of cross-entropy and the Intersection Over Union. This cost function optimizes the overlap between the detected mask and ground-truth while maximizing the probability for a pixel to be correctly classified, previously shown to improve segmentation accuracy (21). The trained CNN models were applied to both stress and rest test scans.

*Training and hyperparameter search*

The data for training was split into a training set (87.5% of all studies) and a validation set (12.5% of all studies) and the CNN model and optimization was implemented using PyTorch (22). Training was performed on a Linux PC (Ubuntu 18.04) with four NVIDIA GTX 2080Ti GPU cards. ADAM optimization was used with initial learning rate being 0.001 (betas are 0.9 and 0.999, epsilon was 1e-8). Learning rate was reduced by x2 for every 10 epochs. Training took 60 epochs and best model was selected as the one giving best performance on the validation set.

*Automated reporting and inline scanner integration*

The trained model was integrated to run on MR scanners using the Gadgetron Inline AI (13) streaming software which provides flexible interfaces to load pre-trained neural networks and apply them on incoming new data. This involved transferring model objects from Pytorch to C++ and passed data from C++ to Pytorch modules. Model inference was chosen to utilize CPU which was sufficiently fast for clinical usage.

Perfusion segmentation functionality was performed after inline perfusion mapping. As soon as a perfusion scan was configured, the pre-trained model was loaded into the Gadgetron runtime environment. Following image reconstruction and pre-processing, models were applied to the incoming 2D+T image series for each slice. Resulting segmentation was used to generate the 16-sector measurement of perfusion and produce a summary report. All steps were fully automatic without any user interaction. A screenshot (Fig. 3) illustrates the perfusion mapping with overlaid CNN based segmentation and AHA report, applied to a patient with reduced regional perfusion. This is a "one-click" solution for automated analysis of quantitative perfusion flow mapping.

*Evaluation of model performance*

The segmentation of automated processing was compared to the manually labeled test set. Performance was quantified in both segmentation accuracy and myocardial flow measures. The Dice ratio (23) for manual label and automatic segmentation masks, was computed, together with the false positive (FP) and false negative (FN) errors. FP was defined as the percentage area of the segmented mask in the CNN result that was not labeled in the manual one. FN was defined as the percentage area of segmented mask in the manual that was not labeled in the automated result. The precision (defined as the percentage of segmented area in both the CNN and manual masks over CNN area) and recall (defined as the percentage of segmented area in both the CNN and manual masks over manual area) were also reported. The myocardium boundary errors (MBE) (24), defined as the mean distance between myocardial borders of two masks, and the Hausdorff distance (25) were computed for the endo- and epicardium borders. The detection accuracy of RV insertion was measured by the angular difference between auto and manual determined direction vectors for RV insertion, as only the orientation was needed for segmentation. Global and per-sector myocardial flow measures were used quantify the CNN performance compared to manual results, displayed using Bland-Altman plots. Additionally, contours were visually inspected for segmentation failures on all 200 test cases.

Results were presented as mean +/- standard deviation. T-test was performed and a P-value less than 0.05 was considered statistically significant.

# Results

An example of segmentation (Fig. 4) illustrates the contours overlaid on perfusion images and corresponding flow maps. The trained CNN correctly delineated the LV cavity and myocardium.

The RV insertion direction was accurately detected to allow sector division. The epicardial fat, was correctly excluded from segmentation and papillary muscles were avoided.

Hyperparameter search was conducted to test different network parameter combinations (2 to 4 resolution layers, 2 to 4 blocks per layer, number of convolution filters either 64, 128 and 256). For all tests, the ADAM optimization was used with initial learning rate being 0.001. The betas were 0.9 and 0.999. Epsilon was 1e-8. After the hyperparameter search, best performance was found for an architecture containing two down-sampling and up-sampling layers, with two ResNet blocks for the first layer and three blocks for the second. This led to a deep net of 23 CONVs in total. On the tested hardware, training took ~8 hours for 60 epochs.

Mean Dice ratio of myocardium segmentation between CNN and manual ground-truth was $0.93 \pm 0.04$. FP and FN were $0.09 \pm 0.06$ and $0.06 \pm 0.05$. Precision and recall were $0.92 \pm 0.06$ and $0.94 \pm 0.05$. MBE was $0.33 \pm 0.15$mm. Given the training image spatial resolution of 1mm$^2$, mean boundary error was less than 1/2 pixel. The mean bidirectional Hausdoff distance was $2.52 \pm 1.08$mm. Mean angle between auto and manually determined RVI directions was $2.65 \pm 3.89$ degree.

Mean stress flow was $2.25 \pm 0.59$ ml/min/g for CNN and $2.24 \pm 0.59$ ml/min/g for manual (P=0.94). For rest scans, CNN gave $1.08 \pm 0.23$ ml/min/g and manual measure gave $1.07 \pm 0.23$ ml/min/g (P=0.83). The per-sector measures showed no significant difference (P=0.92). Bland-Altman plots (Fig. 5) compare auto vs. manual processing of MBF for both global MBF and 16-sector values.

The performance was further evaluated separately for 3T and 1.5T test scans. The mean Dice ratio was $0.93 \pm 0.04$ for 3T and $0.93 \pm 0.03$ for 1.5T (P=0.97). At 3T, the mean stress flow was $2.20 \pm 0.59$ ml/min/g for CNN and $2.21 \pm 0.59$ ml/min/g for manual (P=0.93). The mean rest flow was 1.08 ± 0.23 ml/min/g for CNN and 1.07 ± 0.23 ml/min/g for manual (P=0.84). At 1.5T, the mean stress flow was 2.29 ± 0.60 ml/min/g for CNN and 2.29 ± 0.59 ml/min/g for manual (P=0.97). The mean rest flow was 1.08 ± 0.23 ml/min/g for CNN and 1.07 ± 0.23 ml/min/g for manual (P=0.93).

Contours were visually evaluated on all 200 test cases (3 slices each). There was a single stress case where 1 slice failed to properly segment the RV, however, the myocardium was properly segmented. A second rest case had 1 apical slice where the myocardium segmentation included blood pool. There was apparent through-plane motion that was uncorrected. No other segmentation failures were found.

The CNN model was integrated on the MR scanner. Model loading time was ~120ms and applying model on incomings perfusion series was ~250ms per slice. For a typical three SAX acquisition, inline analysis took <1s on CPU. This timing was measured with CPU inference (2×Intel Xeon Gold 6152 CPU @ 2.1GHz, released in 2017, 192GB RAM). The peak memory usage when applying trained CNN models was ~270MB. On another computer with older hardware (Intel Xeon E5-2680 CPU, released in 2012 and 64G RAM), model loading time was ~130ms and applying models took 370ms per slice.

## Discussion

This paper presents a deep neural network based workflow for automated myocardial segmentation and reporting of the AHA 16 sector model for pixel-wise perfusion mapping. The derived myocardial measures were computed and reported inline on the MR scanner taking just one additional second of inline processing time. Quantitative evaluation in this initial study demonstrated performance of myocardial segmentation and sector-based analysis that is well matched to a human expert. This study used stress and rest data from 7 scanners at 3 sites at 2 field strengths using over 1800 consecutive scans for training and 200 for test. Bland-Altman analysis demonstrated a 95% confidence interval for global MBF of 0.05 mmol/min/g compared to manual labeling, which is sufficient for automated detection and reporting. Prior work on segmenting perfusion (26) used much smaller datasets resulting in much higher variance. A weighted sum loss function was used in this study and gave good accuracy. There are indeed many other alternatives, such as soft Dice ratio or Focal loss (27), which can be effective in perfusion segmentation task. Which loss function is the best may vary for different applications. A comprehensive overview and implementation of many loss functions can be found at *https://github.com/JunMa11/SegLoss*.

The prognostic significance of proposed AI application was studied and recently published in (2). In this study, 1,049 patients went through stress MR perfusion scans and analyzed with proposed CNN models. The finding is "in patients with known or suspected coronary artery disease, reduced MBF and MPR measured automatically inline using artificial intelligence quantification of CMR perfusion mapping provides a strong, independent predictor of adverse cardiovascular outcomes" (2). Thus study demonstrates the relevance of automated myocardial segmentation in CMR stress perfusion. This application was deployed to MR scanners and integrated seamlessly into perfusion imaging. CNN models were applied to perfusion images without any user interaction. Perfusion flow maps and reports were sent to the Dicom database and available for evaluation shortly after image acquisition. No manual processing or extra effort was needed to perform analysis resulting in a time savings for clinicians.

Our approach utilized the temporal information through the whole bolus passage (2D-T) to exploit the contrast dynamics for detecting both RV and LV which enabled finding the RV insertion point. The epicardial fat, showing no dynamic intensity changes, was correctly excluded from segmentation which would be more difficult to avoid on a single static image. Manual labeling was done on a single time frame since respiratory motion correction was employed as a preprocessing step. Conversion of image intensities to units of contrast concentration maintained a consistent dynamic range.

*Limitations*

First, the presented study was conducted on MR scanners from a single vendor. Although the specific myocardial perfusion imaging dual-sequence used may not be available on other platforms, the proposed myocardial segmentation method and CNN models may be still applicable. To encourage researchers on other platforms to adopt the proposed solution, the CNN model files and other resources are shared openly (*https://github.com/xueh2/QPerf*). Second, although the proposed algorithm works well for the vast majority of cases, there have been a few cases that are challenging. For example, in the case of severe hypertrophy some slices may not exhibit any blood pool, i.e., complete extinction. In this case, no endocardial contours are drawn. Third, the CNN models are currently trained for short-axis slices and cannot be applied to long-axis views. New training and test datasets are needed to extend segmentation to long-axis slices. In the case where the slice prescription for short-axis perfusion was imperfect, the basal slice may cover some portion of out-flow tract. In this instance, the proposed algorithm will avoid the blood pool and divide a segment accordingly or may skip a segment entirely. This will result in incomplete segmentation of myocardium. Fourth, in cases of severe respiratory motion that is beyond the range that can be corrected with in-plane retrospective methods, portions of the myocardium may be blurred and CNN segmentation can perform poorly in those region. However, in these cases manual segmentation is difficult as well. Fifth, this study did not involve multiple operators for data labeling. The proposed solution has been deployed to MR scanners and is being further tested in on-going clinical studies. Other myocardial segmentation studies (28) showed good inter-operator agreement on myocardial segmentation on short axis slices, with more discrepancy on LV outflow tract slices.

*Conclusion*

In this paper, we demonstrated automated analysis can be achieved on clinical scanners for perfusion MRI. Image analysis has conventionally been performed on an offline workstation with nontrivial (even tedious) user interaction by clinicians. Deep learning enabled inline analysis immediately after data acquisition as part of imaging computation, therefore more objective, convenient, and faster, reducing clinical burden.

**Abbreviations**

| | |
|---|---|
| AIF | arterial input function |
| AUC | area-under-curve |
| DSC | Dice similarity coefficient |
| FA | flip angle |
| FLASH | fast low angle shot |
| FOV | field-of-view |
| IoU | Intersection over Union |
| MBF | myocardial blood flow |
| MOCO | motion correction |
| PD | proton density |
| SNR | signal-to-noise ratio |
| SR | saturation recovery |
| SSFP | steady state free precession |
| TD | saturation recovery delay time |
| TS | saturation time |
| CNN | convolutional neural network |

**Availability of data and material**

The raw data that support the findings of this study are available from the corresponding author upon reasonable request subject to restriction on use by the Office of Human Subjects Research. The source file to train the CNN model and example datasets are shared at https://github.com/xueh2/QPerf.git.

**Funding**

Supported by the National Heart, Lung and Blood Institute, National Institutes of Health by the Division of Intramural Research and the British Heart Foundation (CH/16/2/32089).

**Authors' contributions**

-- and -- conceived of the study and drafted the manuscript. -- and -- developed the algorithms, implemented the inline integration of neural net model and performed processing and analysis. --, --, --, --, --, --, -- performed the patient studies used to acquire training data. All authors participated in revising the manuscript and read and approved the final manuscript.

# List of Captions

**Figure 1** Data preparation for performing convolutional neural network based segmentation used in this study. Respiratory motion correction of perfusion images provides pixel-wise alignment of myocardial tissue. Image intensities are corrected for surface coil inhomogeneity and converted to Gd concentration units. Images are resampled to a fixed temporal and spatial resolution and cropped around the left ventricular cavity. The resulting 2D+T time series of images is input for CNN training, together with supplied manual labeling.

**Figure 2** Schematic plot of the convolutional neural network trained in this study. This network consists of downsampling and upsampling layers. Each layer includes a number of ResNet blocks. More layers and blocks can be inserted into the CNN to increase its depth. In the example illustration, two layers are used with two blocks for each layer. The total number of convolution blocks is 23.

**Figure 3** Example screen snapshot for a patient undergoing an adenosine stress study, demonstrating the proposed inline analysis solution on a MR scanner. Stress maps show regional flow reduction in septal and inferior sectors. The determined RV insertion was used to split myocardium to AHA sectors, with the contours overlaid to mark territories. The inline reporting further produced a 16-sector AHA bulls-eye plot with global and per-sector flow measures reported in a table.

**Figure 4** Example adenosine stress perfusion images and MBF maps illustrating segmentation in the format of derived AHA sector contours overlaid on flow maps. For each case, the first row are the images in Gd units and the second row are the MBF maps. Sector contours were overlaid to mark three territories for LAD (yellow), RCA (green), and LCX (red). (a) patient with single vessel obstructive CAD in RCA territory. Papillary muscle was not included in segmentation, (b) patient with hypertrophic cardiomyopathy illustrating that the convolutional neural network based segmentation works with thick myocardium and small cavity. The epicardial fat was correctly excluded.

**Figure 5** Bland-Altman plots for independent test dataset (a) global mean myocardial blood flow and (b) per-sector measures. No significant differences were found between CNN derived results and manual measures. The dotted lines mark the 95% confidence range.

Figure 1

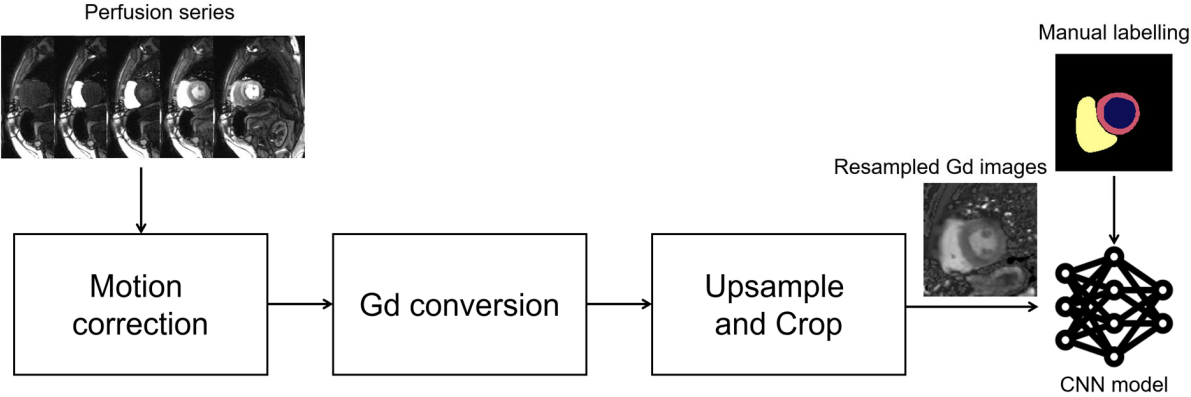

Figure 2

Figure 3

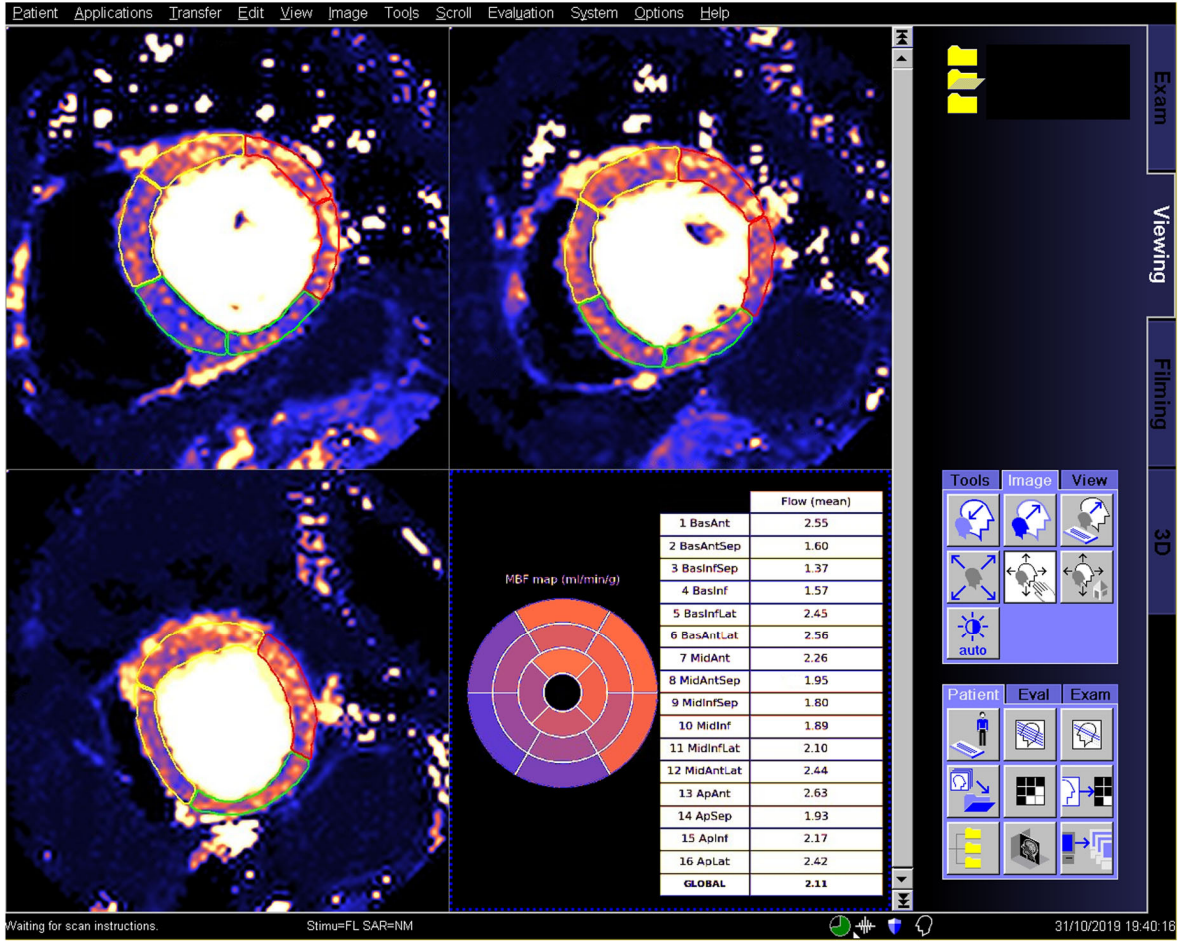

Figure 4

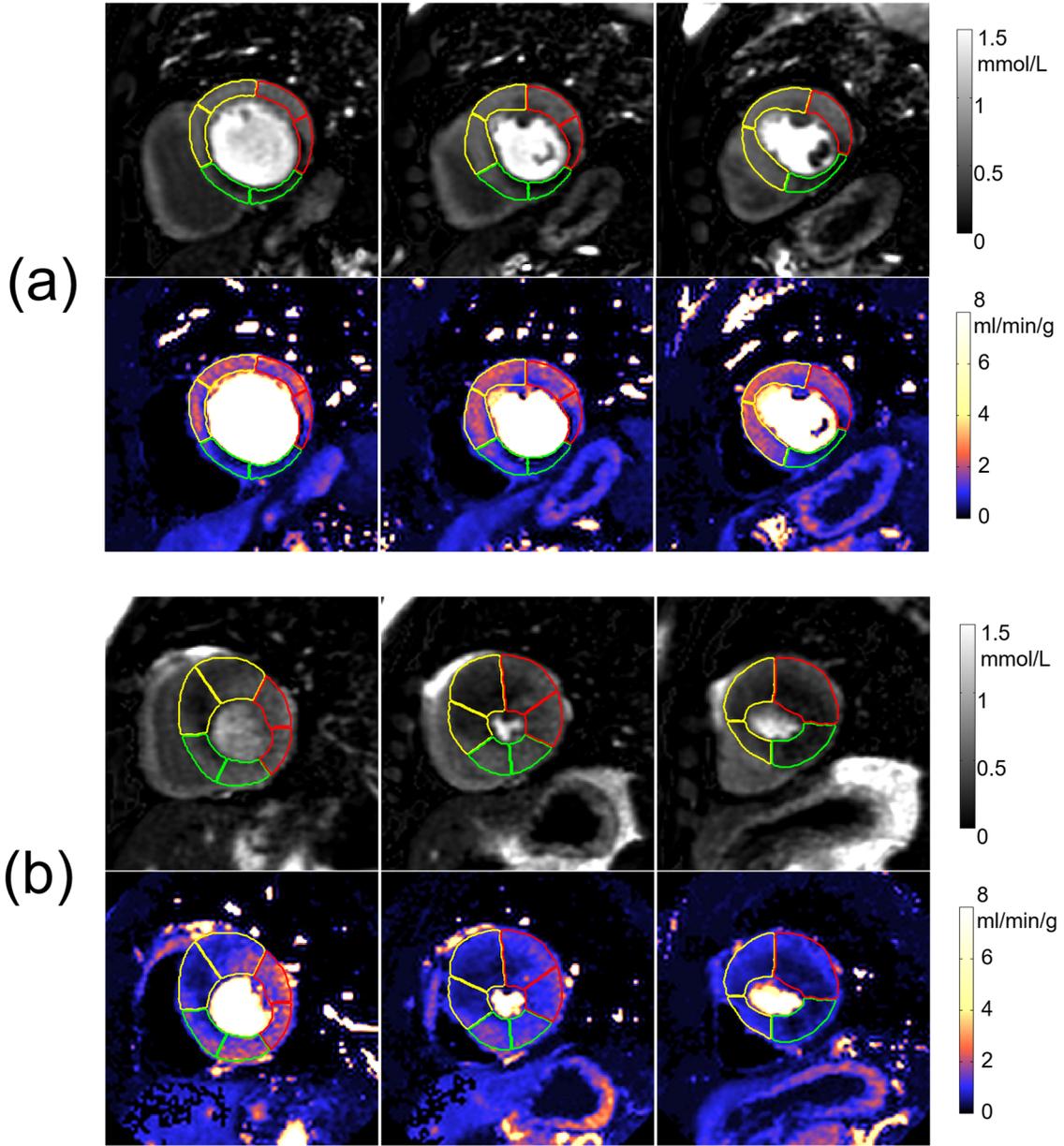

Figure 5

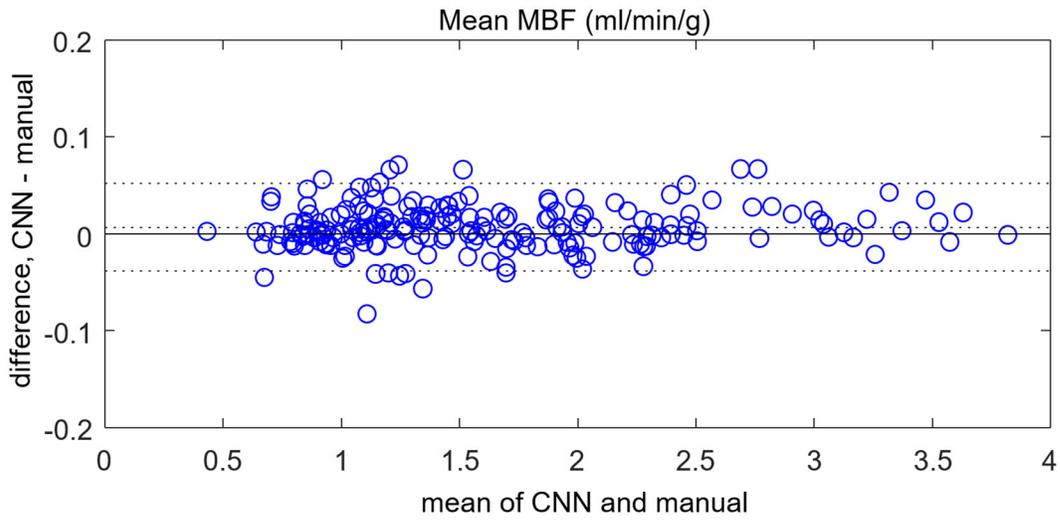

(a) Bland-Altman plot of global MBF

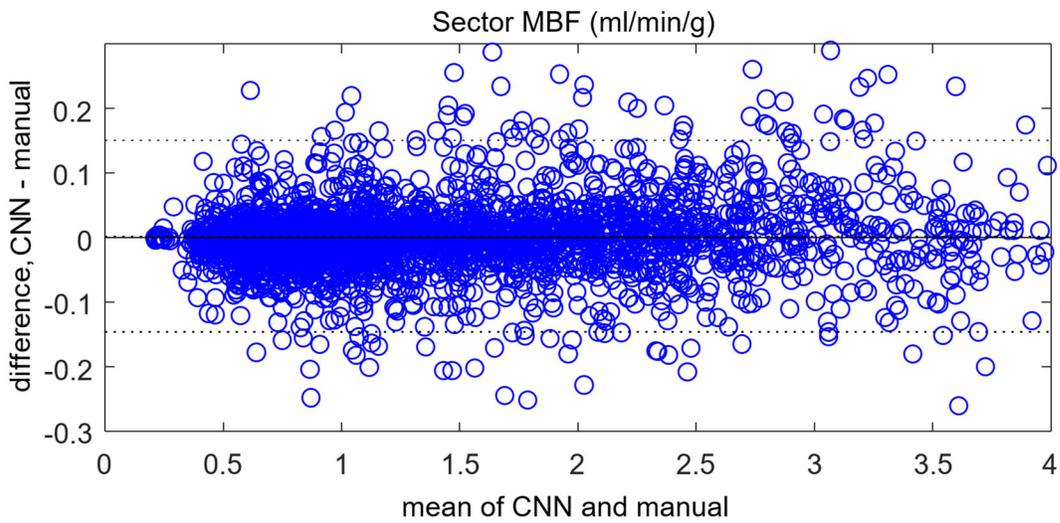

(b) Bland-Altman plot of per-sector MBF